\documentclass[iop]{emulateapj}
\usepackage{url}
\usepackage[utf8]{inputenc}
\usepackage{fancyref}
\usepackage{hyperref}
\hypersetup{colorlinks,breaklinks,
            urlcolor=[rgb]{0.2,0.2,0.2},  
            linkcolor=[rgb]{0.843,0.098,0.110},  
            citecolor=[rgb]{0,0.4,1.0}}  
\usepackage{amssymb,amsmath}
\usepackage{enumitem}
\usepackage{natbib}
\usepackage{epstopdf}
\usepackage{rotating}
\usepackage{color}
\usepackage{float}
\restylefloat{table}
\usepackage{graphicx} 

\newcommand{\teff}{${T}_{\mathrm{eff}}$}
\newcommand{\logg}{$\log{g}$}
\newcommand{\msun}{$M_{\odot}$}

\newcommand{\kms}{km s$^{-1}$}
\newcommand{\vsini}{$v \sin i$}
\newcommand{\muhz}{$\mu$Hz}

\newcommand{\tar}{PG\,0112+104}
\newcommand{\pg}{PG\,1159$-$035}
\newcommand{\kep}{{\em Kepler}}

\newcommand{\ktwo}{{\em K2}}

\newcommand{\bvf}{Brunt-V\"{a}is\"{a}l\"{a} }

\shorttitle{A Deep Test of Internal White Dwarf Rotation}
\shortauthors{Hermes et al.}

\begin{document}

\title{A DEEP TEST OF RADIAL DIFFERENTIAL ROTATION IN A HELIUM-ATMOSPHERE WHITE DWARF: \\I. DISCOVERY OF PULSATIONS IN PG 0112+104}
\author{J. J. Hermes\altaffilmark{1,2}, Steven~D.~Kawaler\altaffilmark{3}, A.~Bischoff-Kim\altaffilmark{4}, J.~L.~Provencal\altaffilmark{5}, B.~H.~Dunlap\altaffilmark{1}, and J.~C.~Clemens\altaffilmark{1} }

\altaffiltext{1}{Department of Physics and Astronomy, University of North Carolina, Chapel Hill, NC 27599-3255, USA}
\altaffiltext{2}{Hubble Fellow}
\altaffiltext{3}{Department of Physics and Astronomy, Iowa State University, Ames, IA 50011, USA}
\altaffiltext{4}{Penn State Worthington Scranton, Dunmore, PA 18512, USA}
\altaffiltext{5}{Department of Physics and Astronomy, University of Delaware, Newark, DE 19716, USA}

\email{jjhermes@unc.edu}

\begin{abstract}

We present the detection of non-radial oscillations in a hot, helium-atmosphere white dwarf using 78.7~d of nearly uninterrupted photometry from the {\em Kepler} space telescope. With an effective temperature $>$$30{,}000$~K, \tar\ becomes the hottest helium-atmosphere white dwarf known to pulsate. The rich oscillation spectrum of low-order $g$-modes includes clear patterns of rotational splittings from consecutive sequences of dipole and quadrupole modes, which can be used to probe the rotation rate with depth in this highly evolved stellar remnant. We also measure a surface rotation rate of 10.17404~hr from an apparent spot modulation in the {\em K2} data. With two independent measures of rotation, \tar\ provides a remarkable test of asteroseismic inference.

\end{abstract}


\section{Introduction}

For centuries, we have judged our stellar neighbors by their covers, amassing broad yet superficial knowledge of the stars in our Galaxy by examining only the light emitted from their photospheres.

An opportunity for deeper inquiry has only been available for the past few decades using stellar seismology, allowing us to challenge the theoretical assumptions that underly ever-improving models. These assumptions matter: stellar models are a critical rung on the ladder to our understanding of extragalactic astronomy, from the integrated stellar light of galaxies to constraining the compact remnants that explode as Supernovae Ia and measure out cosmological distances.

Rotation is a significant property of stars and affects physical processes such as convection, diffusion, and the dynamos generating strong magnetic fields, all of which modify stellar evolution. Even when evolution models incorporate rotation, it is often assumed the star rotates as a solid body. Amazingly, we are now in an age where we can use asteroseismology to deeply test for different effects in stars at all stages of evolution --- revealing internal rotation profiles (e.g., \citealt{Aerts15}) and now even internal magnetic fields (e.g., \citealt{Fuller15}).

One of the significant legacies of the \kep\ mission is the measurement of rotation profiles for a wide range of pulsating stars, from several along the upper main sequence (e.g., \citealt{Kurtz14,Saio15,Murphy16}) to thousands along the red-giant branch (see, e.g., \citealt{Beck12,DiMauro16}). Especially significant is the ability to connect core rotation rates to those in the envelope, probing radial differential rotation and angular momentum evolution. For example, \kep\ data show that the cores of many red giants spin faster than their envelopes (e.g., \citealt{Mosser12,Deheuvels12}). The relatively slow core rates ($\sim$10~d) suggest there is an unknown process efficient at transporting internal angular momentum in evolved stars (e.g., \citealt{Cantiello14,Fuller14}).

\begin{figure*}[t]
\centering{\includegraphics[width=0.995\textwidth]{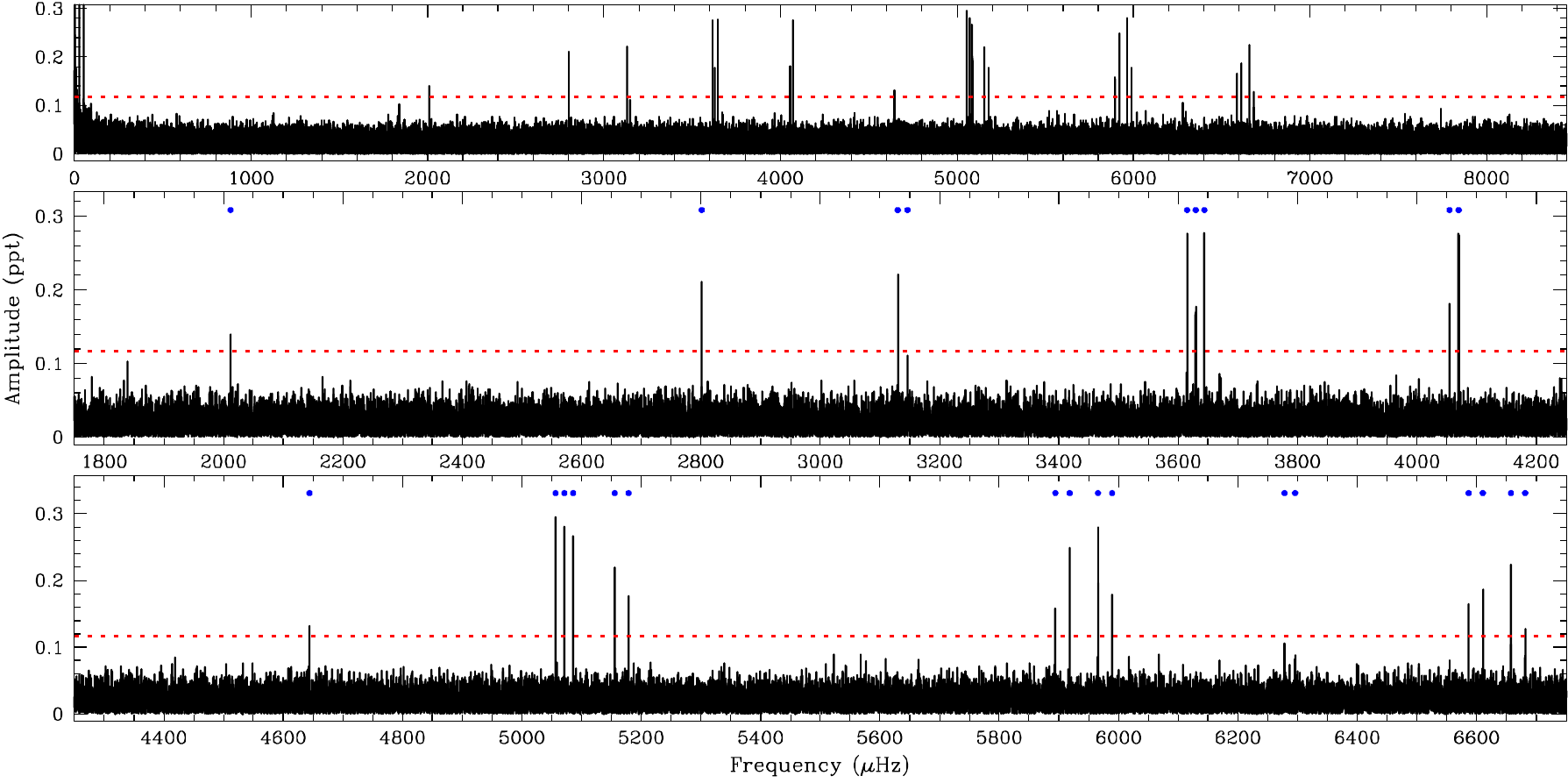}}
\caption{Fourier transforms of the {\em K2} data of \tar. The top panel shows the FT out to the Nyquist frequency of our sampling rate of 58.8 s, and subsequent panels show the regions of pulsation variability in more detail. Figure~\ref{fig:ftrot} shows the low-frequency variability. Our significance threshold is marked as a dotted red line and defined in the text; significant periodicities are marked as blue dots. \label{fig:ft}}
\end{figure*}

There are few empirical constraints on how angular momentum is conserved all the way through the final stages of stellar evolution. Although the prospects have been discussed for nearly two decades \citep{Kawaler99}, only hot pre-white-dwarfs have been studied in sufficient detail to deeply probe their internal rotation profiles, including \pg\ and PG\,0122+200 \citep{Charpinet09,Corsico11,Fontaine13}. Recent efforts have probed less deeply the cooler ZZ Ceti stars (e.g., \citealt{Giammichele16a}).

We present here an exceptional laboratory to connect surface and internal rotation of a highly evolved, apparently isolated star that is now passively losing its residual thermal energy. The white dwarf \tar\ ($g=15.2$~mag, J011437.63+104104.3) was first identified as a helium-atmosphere (DB) white dwarf with strong He I features by \citet{Greenstein77}. It was immediately interesting because it was the hottest DB known before the Sloan Digital Sky Survey filled in the so-called ``DB gap'' \citep{Eisenstein06}. The recent detailed study by \citet{Dufour10} found that \tar\ has \teff$=31300\pm500$ K, \logg$=7.8\pm0.1$ ($M_{\rm WD}=0.52\pm0.05$~\msun).

Several groups have searched for photometric variability in \tar, given its proximity to the DBV (V777~Her) instability strip. \citet{Robinson83} did not see variability greater than 2.1 parts per thousand (ppt) in white-light observations over one night, using a two-channel photometer on the 2.1m at McDonald Observatory. Subsequently, \citet{Kawaler94} saw no significant variability to a limit of 3.1~ppt using roughly 30~min of data taken with the High Speed Photometer on board the {\em Hubble Space Telescope}, using 10~ms exposures through the F140LP ultraviolet filter.

However, two significant periodicities were detected after a five-day campaign in 2001~October in a three-channel photometer on the 2.1m at McDonald Observatory \citep{Shipman02}. Using 30~hr of white-light photometry, \citet{Provencal03} reported variability at $197.76\pm0.01$~s (0.87~ppt) and $168.97\pm0.01$~s (0.83~ppt) not seen in a nearby comparison star. These periods are consistent with the $g$-mode pulsations typically seen in DBVs. Follow-up observations taken over 1.0~hr with a three-channel photometer on the 3.6m Canada-France-Hawaii Telescope in 2002~July could not confirm variability \citep{Dufour10}.

We demonstrate here that \tar\ indeed pulsates at the same periods measured by \citet{Provencal03}, plus nine additional independent pulsation modes. The amplitude of the pulsations are so low that previous non-detections were simply not sensitive enough to observe the variability. We identify five low-order dipole modes and three quadrupole modes, all of which show rotationally split multiplets and probe the internal rotation at different depths. Additionally, we observe a clear photometric signature of the surface rotation rate, likely caused by a spot rotating into and out of view.

\section{Space-Based Time-Series Photometry}
\label{sec:photobs}

We targeted \tar\ for observations during Campaign~8 of the \ktwo\ mission; short-cadence data were obtained by Guest Observer programs in Cycle 3 proposed by Kawaler (GO8011), Hermes (GO8018), and Redfield (GO8048). Cataloged as EPIC~220670436 ($K_p=15.5$~mag), \tar\ was observed beginning 2016~January~04 for 78.723~d with a duty cycle of 96.0\%.

We obtained the short-cadence Target Pixel File (data release version 11) from the Mikulski Archive for Space Telescopes and processed it using the {\sc PyKE} software package managed by the \ktwo\ Guest Observer office \citep{Still12}. We first defined the best pixel extraction using a range of options, and settled on a large aperture with a total of 25~pixels centered around our target. We extracted the light curve, and noticed improved results by not subtracting the median pixel value from each exposure as an estimate of the background. We subsequently fit out a second-order polynomial to flatten the longest-term trends. We then used the {\sc kepsff} task \citep{VJ14} to mitigate motion-correlated noise caused by the regular thruster firings that control the \kep\ roll angle.

We fit out a sixth-order polynomial to this self-flat-fielded light curve to remove long-term thermal systematics in the \ktwo\ light curve. Finally, we iteratively clipped all points falling 3$\sigma$ from the mean using the {\sc VARTOOLS} software package \citep{Hartman16}. Most data were clipped in a region when the spacecraft dropped out of fine point control, leading to a 1.3-d gap beginning roughly 28.1~d from the start of observations. A total of 4591 points were clipped, leaving $110{,}997$ observations. 

\begin{figure}
\vspace{6pt}
\centering{\includegraphics[width=0.995\columnwidth]{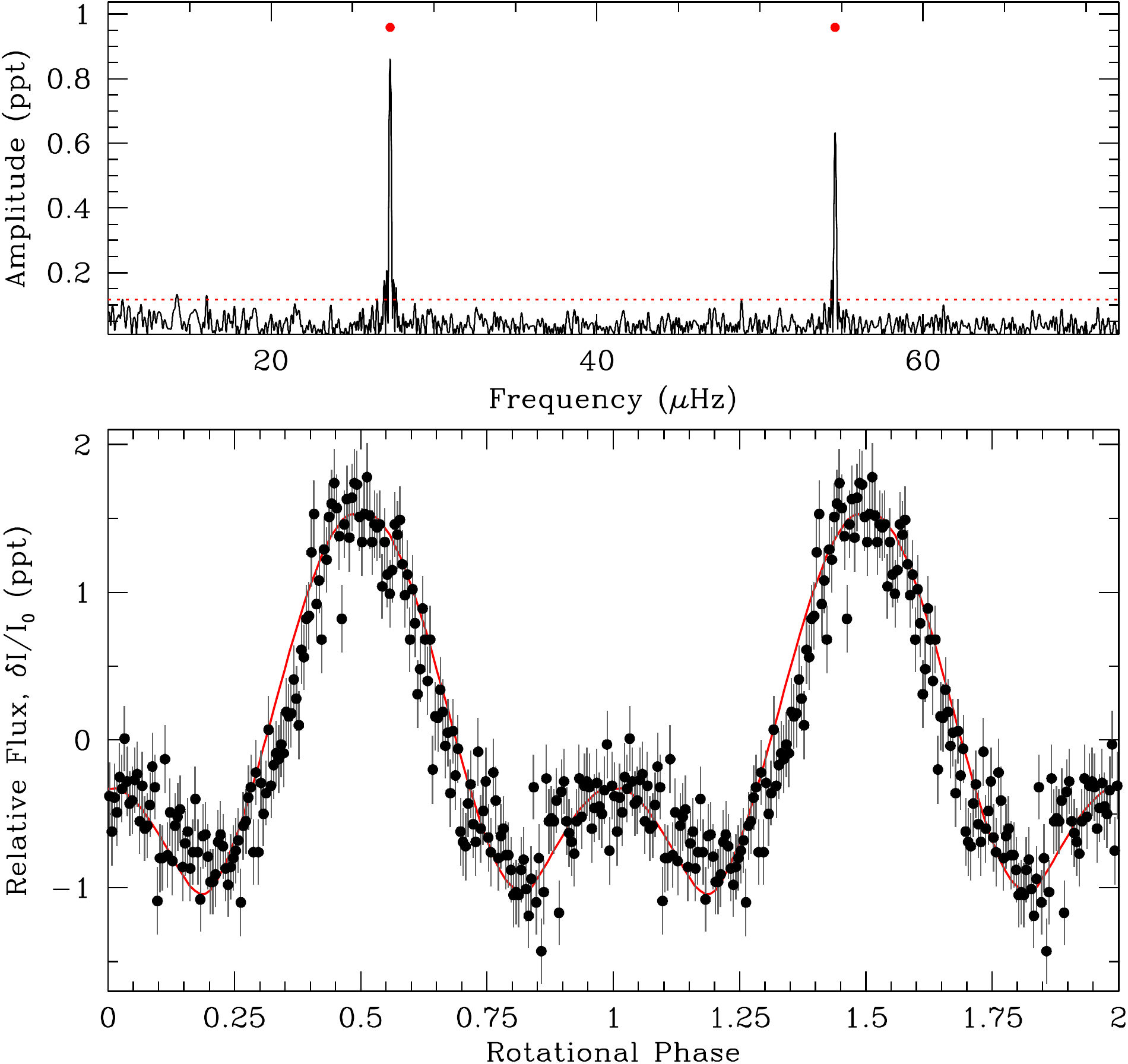}}
\caption{The top panel shows the low-frequency FT of \tar. We identify two significant peaks, marked with red dots, which describe photometric modulation at the rotation period and its first harmonic. The bottom panel shows the {\em K2} light curve binned into 200 points, folded at the rotation period of 10.17404\,hr, and repeated for clarity. A simple spot model is underplotted in red and described in the text.\label{fig:ftrot}}
\end{figure}

The point-to-point scatter in the light curve prevents by-eye detection of the pulsations. However, the signals come clearly into focus after taking a discrete Fourier transform (FT) of the data, which we present in Figure~\ref{fig:ft}. Our FTs here have been oversampled by a factor of 20.

To compute a significance threshold for each peak in the FT we performed a Monte Carlo simulation of the data, as described in \citet{Hermes15}. In short, we keep the time sampling of the data but randomly shuffle the fluxes, creating $10{,}000$ synthetic light curves; we catalog the maximum amplitude at any frequency from the FT of each synthetic light curve. We find that 99.0\% of these synthetic FTs do not have a peak exceeding 0.117~ppt, and 99.7\% do not exceed 0.121~ppt (all synthetic FTs have a maximum peak larger than 0.090~ppt). We set our significance threshold as the value for which there is a $<$1\% chance the signal could arise from noise: 0.117~ppt. This is marked as a red dotted line in all figures. This is nearly identical to 5$\langle {\rm A}\rangle$, where $\langle {\rm A}\rangle$ is the average amplitude of the FT: $5\langle {\rm A}\rangle=0.118$~ppt.

Table~\ref{tab:freq} catalogs all significant frequencies detected in \tar, which were computed using a simultaneous non-linear least squares fit for the frequency, amplitude, and phase of all peaks using the software {\sc PERIOD04} \citep{Lenz05}. The high precision of the \kep\ data actually requires a frequency correction due to the gravitational redshift and line-of-sight Doppler motion of the white dwarf \citep{Davies14}. Using the measured radial velocity of $34\pm3$~\kms\ \citep{Provencal00}, we transform the observed frequencies ($\nu_o$) to those in the stellar rest frame ($\nu_s$) with the relationship $\nu_s=0.9998866\nu_o$. All frequencies arising from pulsations and rotation have been corrected to the stellar rest frame in Table~\ref{tab:freq} (the phases were unchanged). All amplitudes from our least-squares fit have a formal uncertainty of 0.019~ppt. The least-squares fit includes the instrumental electronic (long-cadence) artifacts listed in Table~\ref{tab:freq}. Phases are measured relative to the mid-point of the first exposure: 2457392.0589289~BJD$_{\rm TDB}$.

\begin{deluxetable}{lrrcr}
\tablecolumns{5}
\tablewidth{0.45\textwidth}
\tablecaption{Frequencies present in \tar\
  \label{tab:freq}}
\tablehead{\colhead{ID} & \colhead{Frequency} & \colhead{Period} & \colhead{Amplitude} & \colhead{Phase}
\\ \colhead{} & \colhead{($\mu$Hz)} & \colhead{(s)} & \colhead{(ppt)} & \colhead{(rad/2$\pi$)} }
\startdata \\
\multicolumn{5}{c}{{\em Spot Modulation \& Harmonics}} \\ \\
$f_{{\rm spot}}$ & 27.3026(18) & $36{,}626.6$ & 0.864 & 0.9382(36) \\ 
2$f_{{\rm spot}}$ & 54.6052(25) & $18{,}313.3$ & 0.633 & 0.6183(48) \\ 
3$f_{{\rm spot}}$ & 81.908(22) & $12{,}208.9$ & 0.070 & 0.339(44) \\ \\
\hline \\
\multicolumn{5}{c}{{\em Independent Pulsation Modes, Ordered by Amplitude}} \\ \\
$f_{1a}$ & 5055.9298(52) & 197.78756 & 0.298 & 0.225(10) \\ 
$f_{1b}$ & 5070.7957(56) & 197.20771 & 0.281 & 0.953(11) \\ 
\multicolumn{5}{r}{$f_{1b}-f_{1a}=14.866(11)$ \muhz} \\
$f_{1c}$ & 5085.6610(59) & 196.63127 & 0.267 & 0.003(11) \\ 
\multicolumn{5}{r}{$f_{1c}-f_{1b}=14.865(11)$ \muhz} \\
$f_{2a}$ & 5893.4822(99) & 169.67897 & 0.157 & 0.788(19) \\ 
$f_{2b}$ & 5917.3195(63) & 168.99544 & 0.249 & 0.999(12) \\ 
\multicolumn{5}{r}{$f_{2b}-f_{2a}=23.837(16)$ \muhz} \\
$f_{2c}$ & 5964.8940(56) & 167.64757 & 0.281 & 0.062(11) \\ 
$f_{2d}$ & 5988.6106(87) & 166.98364 & 0.179 & 0.456(17) \\ 
\multicolumn{5}{r}{$f_{2d}-f_{2c}=23.717(14)$ \muhz} \\
$f_{3a}$ & 3614.4976(56) & 276.66362 & 0.277 & 0.029(11) \\ 
$f_{3b}$ & 3628.9890(88) & 275.55884 & 0.177 & 0.782(17) \\ 
\multicolumn{5}{r}{$f_{3b}-f_{3a}=14.491(14)$ \muhz} \\
$f_{3c}$ & 3643.4982(56) & 274.46150 & 0.278 & 0.403(11) \\ 
\multicolumn{5}{r}{$f_{3c}-f_{3b}=14.509(14)$ \muhz} \\
$f_{4a}$ & 4054.2233(87) & 246.65637 & 0.179 & 0.998(17) \\ 
$f_{4b}$ & 4069.6229(57) & 245.72301 & 0.275 & 0.698(11) \\ 
\multicolumn{5}{r}{$f_{4b}-f_{4a}=15.400(14)$ \muhz} \\
$f_{5a}$ & 6586.1032(95) & 151.83485 & 0.165 & 0.729(19) \\ 
$f_{5b}$ & 6609.8257(83) & 151.28992 & 0.188 & 0.091(16) \\ 
\multicolumn{5}{r}{$f_{5b}-f_{5a}=23.722(18)$ \muhz} \\
$f_{5c}$ & 6657.1358(70) & 150.21475 & 0.223 & 0.325(14) \\ 
$f_{5d}$ & 6680.749(12) & 149.68382 & 0.127 & 0.121(24) \\ 
\multicolumn{5}{r}{$f_{5d}-f_{5c}=23.613(19)$ \muhz} \\
$f_{6a}$ & 5155.0647(71) & 193.98399 & 0.221 & 0.162(14) \\ 
$f_{6b}$ & 5178.2168(88) & 193.11667 & 0.178 & 0.432(17) \\ 
\multicolumn{5}{r}{$f_{6b}-f_{6a}=23.152(16)$ \muhz} \\
$f_{7a}$ & 3129.7151(71) & 319.51790 & 0.221 & 0.616(14) \\ 
$f_{7b}$ & 3145.892(14) & 317.8748 & 0.111 & 0.281(28) \\ 
\multicolumn{5}{r}{$f_{7b}-f_{7a}=16.177(21)$ \muhz} \\
$f_{8a}$ & 2801.2608(74) & 356.98212 & 0.212 & 0.020(14) \\ 
$f_{9a}$ & 2011.357(11) & 497.1768 & 0.140 & 0.154(22) \\
$f_{10a}$ & 4643.606(12) & 215.34986 & 0.131 & 0.663(23) \\ 
$f_{11a}$ & 6277.488(15) & 159.29938 & 0.105 & 0.833(29) \\ 
$f_{11b}$ & 6295.214(18) & 158.85084 & 0.086 & 0.565(36) \\ 
\multicolumn{5}{r}{$f_{11b}-f_{11a}=17.726(33)$ \muhz} \\ \\
\hline \\
\multicolumn{5}{c}{{\em Instrumental Electronic (Long-Cadence) Artifacts}} \\ \\
6$f_{{\rm LC}}$ & 3398.853(17) & 294.21694 & 0.092 &  \\ 
7$f_{{\rm LC}}$ & 3965.3387(60) & 252.18527 & 0.260 &  \\ 
8$f_{{\rm LC}}$ & 4531.6946(40) & 220.66800 & 0.390 &  \\ 
8$f_{{\rm LC}}$ & 4531.7820(14) & 220.663747 & 1.116 &  \\ 
8$f_{{\rm LC}}$ & 4531.9313(46) & 220.65648 & 0.337 &  \\ 
9$f_{{\rm LC}}$ & 5098.2777(67) & 196.14467 & 0.232 &  \\ 
10$f_{{\rm LC}}$ & 5664.7749(97) & 176.52952 & 0.162 &  \\ 
11$f_{{\rm LC}}$ & 6231.260(11) & 160.48119 & 0.146 &  \\
12$f_{{\rm LC}}$ & 6797.7278(95) & 147.10798 & 0.164 &  \\ 
13$f_{{\rm LC}}$ & 7364.180(10) & 135.79244 & 0.157 &  \\ 
14$f_{{\rm LC}}$ & 7930.6589(65) & 126.09293 & 0.241 &  \\ 
14$f_{{\rm LC}}$ & 7930.818(17) & 126.09040 & 0.093 &  \\ \\
\hline \\
\multicolumn{5}{c}{{\em Frequencies Not Meeting Significance Threshold}} \\ \\
$\omega_{1} $ & 1838.395(15) & 543.9529 & 0.103 & 0.834(30) \\ 
$\omega_{2} $ & 7740.455(17) & 129.19137 & 0.093 & 0.547(33) \\
\enddata
\tablecomments{All periods and frequencies (except instrumental electronic artifacts) have been Doppler shifted to the stellar rest frame; see text for details.}
\end{deluxetable}

\begin{figure}
\centering{\includegraphics[width=0.995\columnwidth]{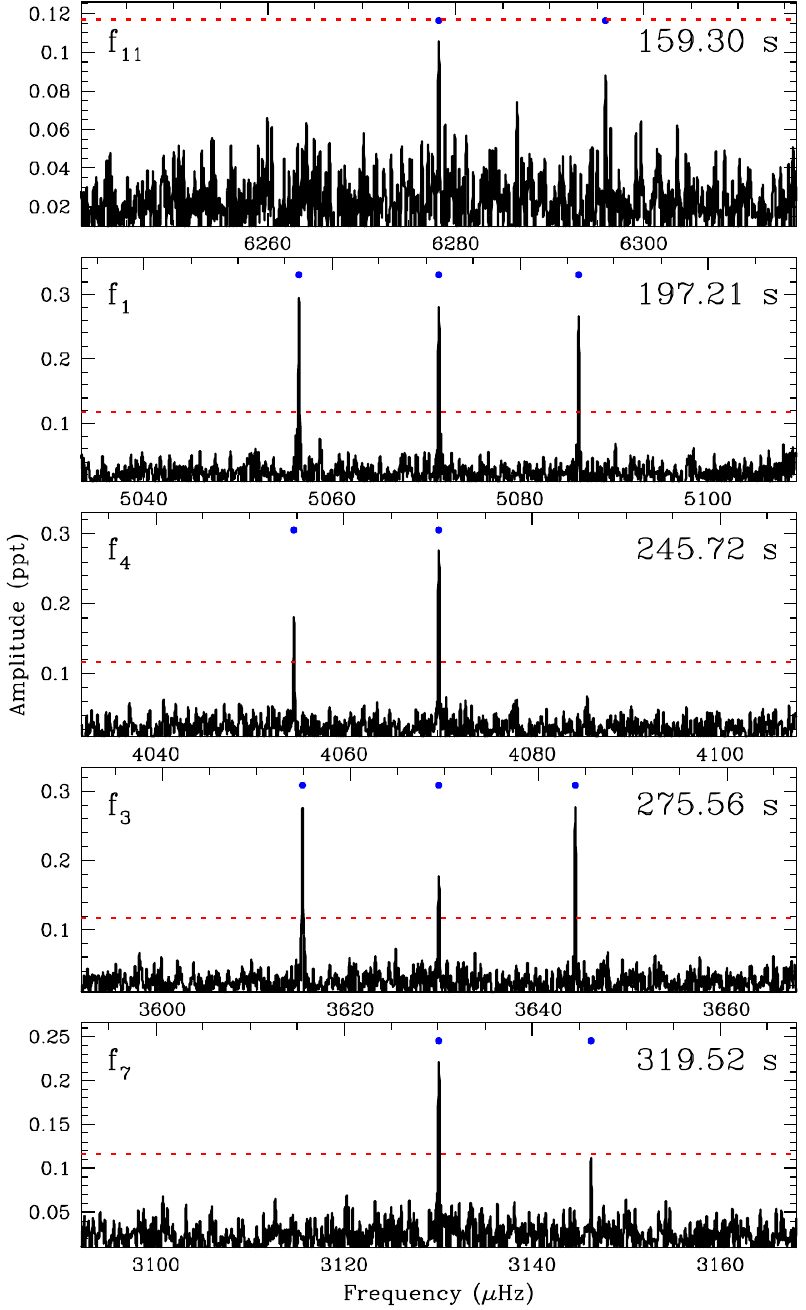}}
\caption{The observed sequence of $\ell=1$ dipole modes present in \tar, plotted in ascending radial order. \label{fig:dipoles}}
\end{figure}

\begin{figure}
\centering{\includegraphics[width=0.995\columnwidth]{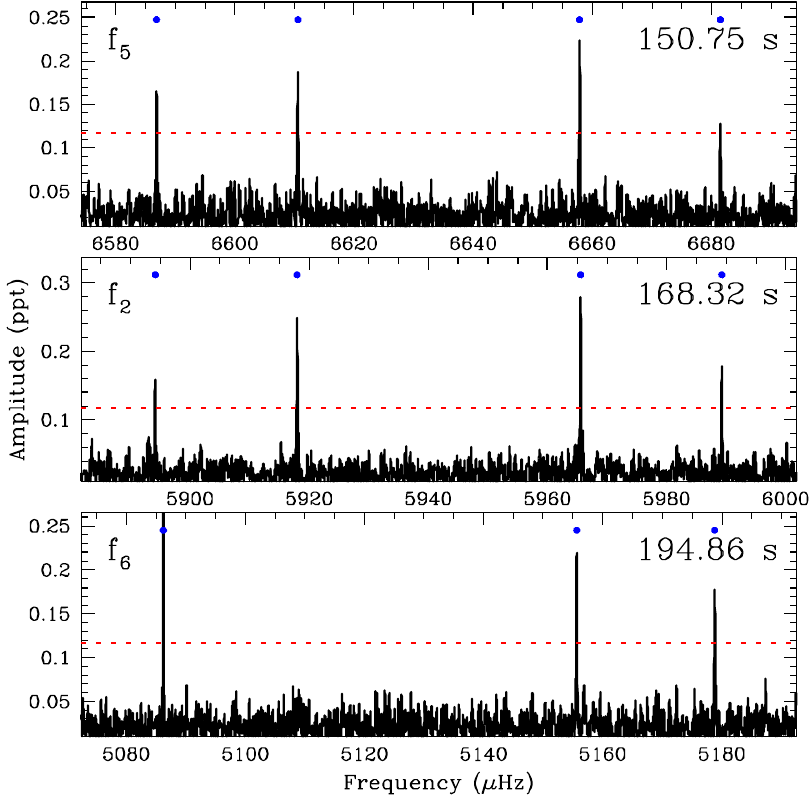}}
\caption{The observed sequence of $\ell=2$ quadrupole modes present in \tar, plotted in ascending radial order. 
We see a pattern consistent with the expected quintuplets, if we allow that inclination effects have suppressed the central $m=0$ components. The lowest-frequency, $m=-2$ component of $f_6$ appears to overlap with the highest-frequency, $m=+1$ component of $f_1$.
\label{fig:quads}}
\end{figure}

\section{Low-Frequency Spot Modulation}
\label{sec:spot}

Immediately evident in the low-frequency FT of \tar\ is a peak at 27.3026~\muhz\ and its first harmonic, shown in more detail in Figure~\ref{fig:ftrot}. The low-frequency variability in Figure~\ref{fig:ftrot} is not related to the 5.88-hr \ktwo\ thruster firing timescale. By folding the entire light curve into 200 phase bins, we see strong evidence for a spot coming into and out of view at what would correspond to a surface rotation period of $10.17404\pm0.00067$\,hr. Further evidence from asteroseismology strengthens this claim (see Section~\ref{sec:pulsations}).

We show a simple two-bright-spot (offset dipole) model in Figure~\ref{fig:ftrot}, using an inclination of 60 degrees with spots at latitude $\pm$34 degrees. This is the first time a photometric rotation rate has been observed in a pulsating white dwarf, and its source is not immediately evident. The most likely cause is the redistribution of flux from the ultraviolet to visible wavelengths, causing an optical bright spot. While \citet{Dufour10} put stringent photospheric upper limits on 12 heavy elements in \tar, several C II and C III absorption features are observed in the ultraviolet. If this carbon is confined to a pole by a relatively small magnetic field, it could serve as an opacity source that rotates into and out of view. This is observed in other white dwarfs. For example, an observed phase difference between the 555\,s X-ray and optical modulation of the $32{,}400$\,K \citep{Barstow97} accreting white dwarf in V471~Tau suggests that metals on the white dwarf line-blanket the high-energy flux, inflating the longer-wavelength brightness at the white dwarf rotation period \citep{Clemens92}. Time-series ultraviolet spectroscopy could confirm this hypothesis in \tar, since the surface abundance of the carbon lines should change with rotational phase.

Regardless of the exact cause of the photometric variability, we measure a surface rotation period of $10.17404\pm0.00067$\,hr in \tar, in line with rotation rates derived from asteroseismology of isolated white dwarfs \citep{Kawaler15}. The spectroscopically derived surface gravity (\logg\ $= 7.8\pm0.1$) and mass ($0.52\pm0.05$ \msun) of \citet{Dufour10} suggest a corresponding \vsini\ $=1.8\pm0.3$ \kms. \citet{Dufour02} showed that \tar\ does not have an abnormally large surface velocity (\vsini\ $<10$~\kms), despite suggestions from an analysis by \citet{Provencal00}.

\section{Observed Pulsation Modes}
\label{sec:pulsations}

There are clear patterns of frequency spacing in the FT of \tar, which point to a lifting of degeneracy in the pulsations caused by rotation; the non-radial $g$-mode oscillations of white dwarfs can be decomposed into spherical harmonics of radial order, $n$, spherical degree, $\ell$, and azimuthal order, $m$ \citep{Unno89}. To first order, the frequency splittings ($\delta \nu$) are related to the stellar rotation ($\Omega$) by the relation
\begin{equation}
\delta \nu = m (1 - C_{n,\ell}) \Omega
\end{equation}
where $C_{n,\ell}$ represents the effect of the Coriolis force on the pulsations as formulated by \citet{Ledoux51}. Figure~\ref{fig:dipoles} shows the $14.5-17.7$\,\muhz\ splittings which correspond to the $\ell=1$ modes present, while Figure~\ref{fig:quads} shows the $23.2-23.8$\,\muhz\ splittings corresponding to $\ell=2$ modes.

Significantly, the observed splittings suggest a white dwarf rotation rate of roughly 10\,hr. Assuming solid-body rotation at 10.17404~hr in the limit such that the horizontal displacement is much larger than the vertical displacement (which is the case for $g$-modes in white dwarfs) --- $C^*_{n,\ell}\approx1/\ell(\ell+1)$ --- implies rotational splittings for $\ell=1$ modes of 13.7\,\muhz\ and 22.8\,\muhz\ for $\ell=2$ modes. Sensitivity to mode-trapping effects from structural discontinuities in the white dwarf model push low-order modes away from the asymptotic limit ($C_{n,\ell}<C^*_{n,\ell}$), so we expect these values to be lower limits for the amount of the rotational splittings \citep{Kawaler99}. This is consistent with what we observe in the frequency differences in Table~\ref{tab:freq}.

All three peaks of the two highest-amplitude dipole modes are significant, unambiguously identifying the central components of $f_1$ as $197.20771\pm0.00022$~s and $f_3$ as $275.55884\pm0.00067$~s. The next two dipole modes only have two significant peaks, complicating mode identification. For $f_4$ we see weak evidence for a peak at $4085.195\pm0.029$~\muhz\ (0.053~ppt), so we propose the $m=0$ component of $f_4$ is $245.72301\pm0.00034$~s. Similarly, we see marginal evidence for a peak at $3112.322\pm0.023$~\muhz\ (0.067~ppt), so we propose the central component of $f_7$ is $319.51790\pm0.00072$~s.

None of the peaks around 6277~\muhz\ are formally significant, but their frequency spacing fits our established pattern. Our alternative light curve extraction calculated from the same aperture but subtracting the background provides stronger support for this as an independent mode. It shows three nearly significant peaks perfectly symmetric about the central component (within the uncertainties): 6259.721(0.018) \muhz\ (0.097~ppt), 6277.488(0.015) \muhz\ (0.119~ppt), and 6295.214(0.015) \muhz\ (0.115~ppt); the significance threshold in that extraction was 0.129 ppt. Therefore, we propose $f_{11}$ is a dipole mode centered at $159.29938\pm0.00038$~s.

\begin{deluxetable}{lrllrlrll}
\tablecolumns{9}
\tablewidth{0.45\textwidth}
\tablecaption{Periods of Likely $m=0$ Components
  \label{tab:m0}}
\tablehead{\colhead{ID} & \colhead{Period} & \colhead{$\ell$} & \colhead{$n$} &  & \colhead{ID} & \colhead{Period} & \colhead{$\ell$} & \colhead{$n$} 
\\  & \colhead{(s)} & & & & \colhead{} & \colhead{(s)} & \colhead{} & }
\startdata
$f_{11}$ & 159.29932 & 1 & 2  &  & $f_{5}$ & 150.7506 & 2 & 4 \\  
$f_{1}$  & 197.20771 & 1 & 3  &  & $f_{2}$ & 168.3185 & 2 & 5 \\ 
$f_{4}$  & 245.72301 & 1 & 4  &  & $f_{6}$ & 194.8591 & 2 & 6 \\
$f_{3}$  & 275.55884 & 1 & 5  & & $f_{10}$ & 216.447 & 2? & 7? \\  
$f_{7}$  & 319.51790 & 1 & 6   &  &   &   &   &  \\ 
$f_{8}$  & 356.98212 & 1? & 7? &  & $f_{9}$ & 497.1768   & 1? &  
\enddata
\end{deluxetable}

\begin{figure}
\centering{\includegraphics[width=0.995\columnwidth]{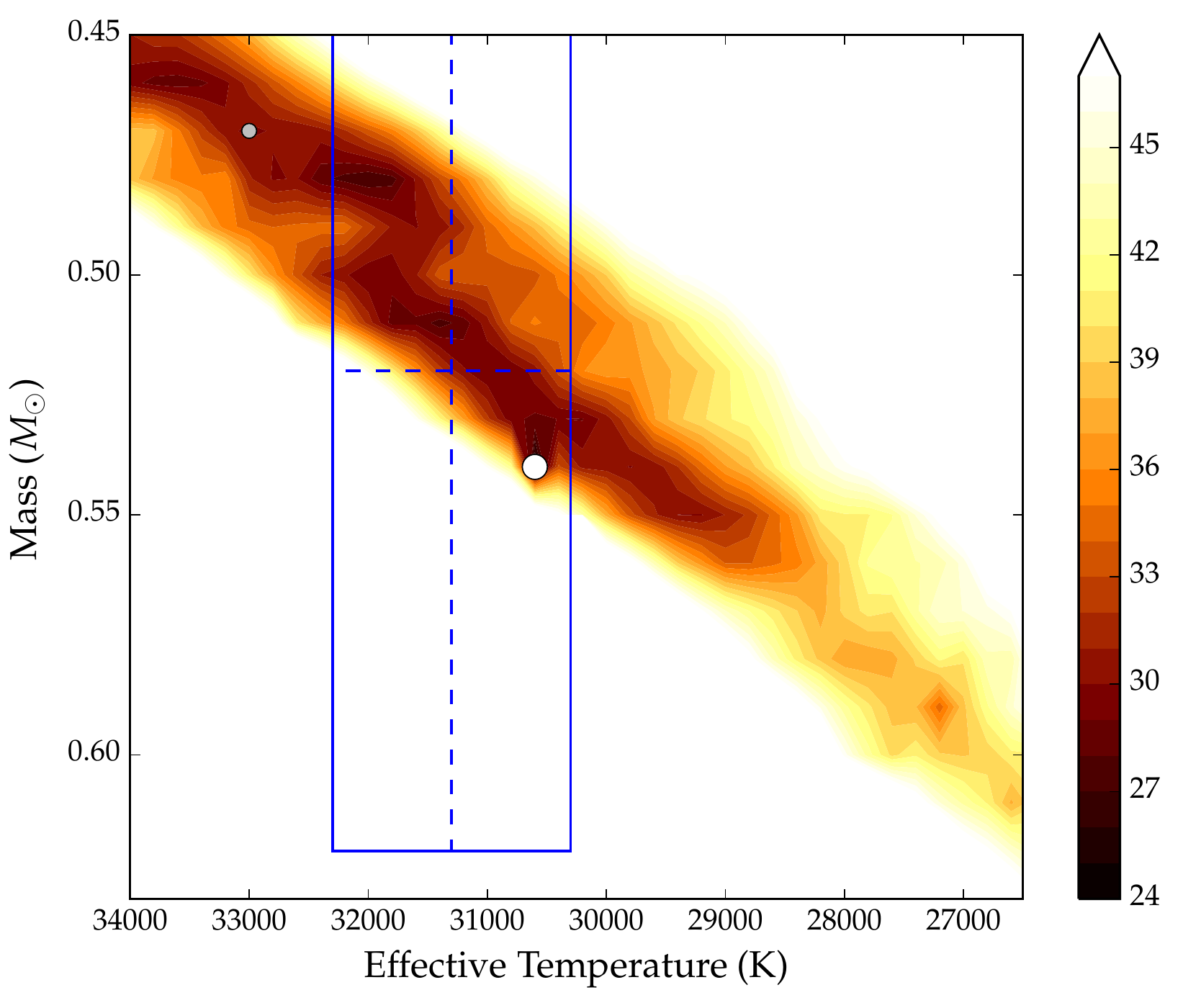}}
\caption{Contours of the goodness of fit, $\sigma_{\rm RMS}$, from our preliminary asteroseismic analysis of \tar. The scale runs linearly in tenths of seconds; darker regions correspond to a better fit, such that the darkest colors visible are twice as good a fit as the lightest colors. We mark the best asteroseismic solution as a white dot, and mark the 2$\sigma$ spectroscopically determined atmospheric parameters from \citet{Dufour10} in blue. The grey point marks the best fit if $f_9$ is an $\ell=2$ mode. \label{fig:contour}}
\end{figure}

We observe three sets of frequencies that appear to correspond to a consecutive sequence of quadrupole ($\ell~=~2$) modes. In two cases, the spacings clearly indicate that the central component is absent. This is likely a geometric effect; the $m=0$ component of an $\ell=2$ mode is strongly suppressed when $50 < i < 60$ deg \citep{Pesnell85,Brassard95,GS03}. We subtract the average frequency difference from the $m=+1$ component to find that $f_5$ is centered at $150.7506\pm0.0010$~s, $f_2$ at $168.3185\pm0.0010$~s, and $f_6$ at $194.8591\pm0.0018$~s. Interestingly, the $m=-2$ component of $\ell=2$ mode $f_6$ appears to overlap identically with the $m=+1$ component of the $\ell=1$ mode $f_1$.

Additionally, we see three significant frequencies that each appear to be an additional independent mode, but we cannot clearly identify them since they lack multiplet splittings. Because there is not likely another $\ell=1$ mode between $f_2$ and $f_3$, which appear to be consecutive overtones, we tentatively identify $f_{10}$ as an $\ell=2$ mode. At inclinations which suppress the $m=0$ component of $\ell=2$ modes, the $m=\pm1$ components are more likely to be observed at higher amplitude than the $m=\pm2$ components (there is generally no preference for multiplet components of $\ell=1$ modes). We tentatively propose the significant peak for $f_{10}$ is the $m=+1$ component (since this is the highest component of all three well-identified $\ell=2$ modes), subtract the mean splitting of the $\ell=2$ modes (23.53~\muhz), and propose $f_{10}$ has a central component at $216.447\pm0.013$~s. (If we are instead observing the $m=-1$ component, $f_{10}$ would be centered at $214.264\pm0.013$~s.)

Since it appears close to the mean period spacing for the $\ell=1$ modes, we propose that $f_8$ is an $\ell=1$ mode. In the absence of further constraints, we propose the $m=0$ component of $f_8$ is $356.98212\pm0.00094$~s. (However, it is equally likely that this mode is centered at $359.03\pm0.17$~s or $354.96\pm0.17$~s, if we are observing the $m=-1$ or $m=+1$ component, respectively, using the 15.99~\muhz\ mean splitting of the other identified $\ell=1$ modes.)

The longest-period mode in the star, $f_9$, has the least-secure identification. Incorrectly assuming its azimuthal order makes the largest difference in period: if it is the $m=+1$ component of an $\ell=2$ mode, $f_9$ would be centered at 503.062~s, a difference of nearly 5.9~s from if it were the $m=0$ component of an $\ell=1$ mode. We tentatively identify $f_{9}$ as a dipole mode centered at $497.1768\pm0.0028$~s (although $501.16\pm0.33$~s or $493.26\pm0.32$~s are equally likely to be the $m=0$ component). We also explore the best fits if $f_{9}$ is an $\ell=2$ mode centered at $503.062\pm0.072$~s.

\section{Preliminary Asteroseismic Analysis}
\label{sec:astero}

The period of the $m=0$ component of each individual mode forms the input for our asteroseismic analysis. We collect these 11 periods and tentative mode identifications in Table~\ref{tab:m0}, although note that the exact $m$ identification is not absolutely secure for modes $f_6-f_{10}$.

We plot our exploration of the minimum goodness-of-fit in Figure~\ref{fig:contour}, where we have used a grid-based approach to match the observed pulsation periods to theoretical ones computed by adiabatic calculations using an updated version of the White Dwarf Evolution Code \citep{LvH75,Wood90}. Further details of our models, methodology, and the goodness-of-fit parameter we minimize, $\sigma_{\rm RMS}$, can be found in \citet{BK14}.

\begin{figure}
\centering{\includegraphics[width=0.995\columnwidth]{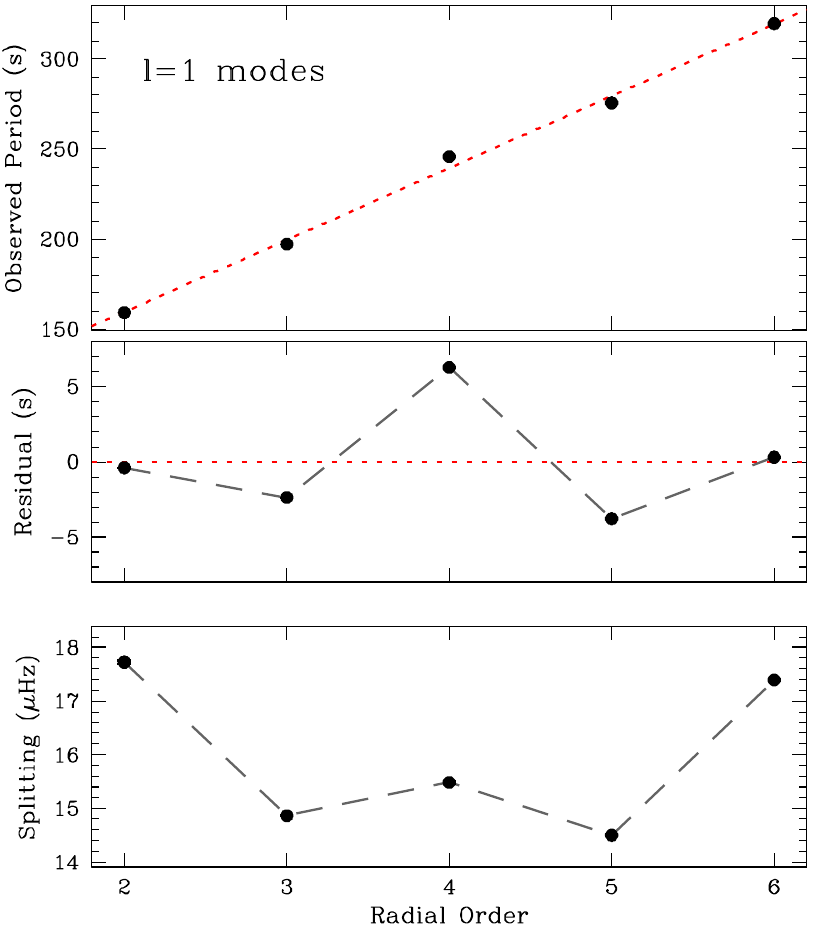}}
\caption{Top: The inferred $m=0$ periods for the consecutive series of $\ell=1$ modes identified in Figure~\ref{fig:dipoles}. The first-order best fit estimates the asymptotic mean period spacing of $39.9\pm2.6$~s. Middle: Deviation from the asymptotic period spacing. Bottom: The average splittings for each radial order, which are correlated with the deviations from the asymptotic period spacing. Both result from mode-trapping effects. \label{fig:splitsdipoles}}
\end{figure}

\begin{figure}
\centering{\includegraphics[width=0.995\columnwidth]{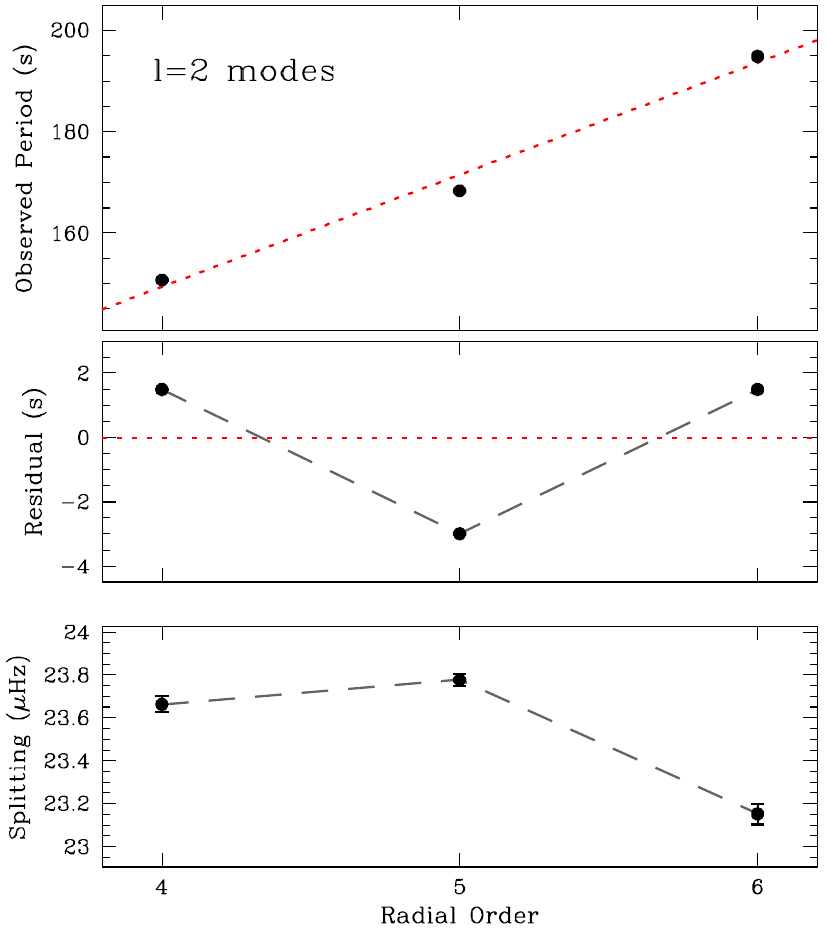}}
\caption{Top: Similar to Figure~\ref{fig:splitsdipoles}, we show the inferred $m=0$ periods for the consecutive series of $\ell=2$ modes identified in Figure~\ref{fig:quads}. The first-order best fit estimates the asymptotic mean period spacing of $22.1\pm2.0$~s. Middle: Deviation from the asymptotic period spacing. Bottom: The average splittings for each radial order, which are correlated with the deviations from the asymptotic period spacing. \label{fig:splitsquads}}
\end{figure}

The models have six variable parameters, which we detail here, including the step size for each grid point: effective temperature ($26{,}200$~K, $34{,}000$~K, $\Delta$200 K), overall mass (0.45~\msun, 0.70~\msun, $\Delta$0.01~\msun), the central oxygen abundance relative to carbon ($X_{\rm O}$; 0.1, 1.0, $\Delta$0.1), the mass fraction of the homogeneous carbon-oxygen core ($X_{\rm fm}$; 0.10, 0.80, $\Delta$0.05), the location of the mixed carbon-helium layer in logarithmic mass coordinates ($q_{\rm env}$; $-3.4$, $-2.0$, $\Delta$0.2 dex), and the location at which the helium abundance rises to 1 ($q_{\rm He}$; $-7.0$, $-4.0$, $\Delta$0.2 dex). Our grid has more than 10 million points.

We inspect just the effective temperature and overall mass, which can be determined via independent methods, in Figure~\ref{fig:contour}. The best-fitting values for the other parameters were optimized at each grid point to make these two-dimensional projections.

There are several ridges of acceptable fits to the periods as identified in Table~\ref{tab:m0}. Our best fit, marked in white at $30{,}600$~K, 0.54~\msun\ in Figure~\ref{fig:contour}, has $\sigma_{\rm RMS}=2.471$~s; the structural parameters for this best-fit are $X_{\rm O}=0.1$, $X_{\rm fm}=0.80$, $q_{\rm env}=-2.6$, and $q_{\rm He}=-4.8$. Unfortunately, two structural parameters are at grid edges, and the residuals of the best-fit are several orders of magnitude larger than the observed period uncertainties. The second-best fit is only marginally worse, at $32{,}000$~K, 0.48~\msun, with $\sigma_{\rm RMS}=2.702$~s. The best fit excluding $f_9$ occurs along the same temperature-mass contour shown in Figure~\ref{fig:contour} but much hotter, at $33{,}800$~K, 0.46~\msun.

We have used this best-fit to assign the radial order of the modes identified in Table~\ref{tab:m0}, since these values are consistent for all best-fit solutions. Our preliminary asteroseismology also consistently points to this DBV being hotter than $30{,}500$~K. However, we do not believe our best fit is confidently at a global minimum.

This is unfortunate: Once we are able to calculate the absolute best-fit to the observed periods, we can then perform an inversion to measure the rotation frequencies at various depths, testing for radial differential rotation. This occurs because modes of different radial order can sample different regions of the star from other modes of the same spherical degree. Some modes are more sensitive to deeper regions in the star, while others are trapped in the outer layers; this allows us to probe the rotation rate with depth. We will undertake this more detailed analysis in a future publication.

The modes are restricted to different regions of the star depending on where the radial nodes coincide with sharp changes in the \bvf\ frequency, which are caused by rapid changes in the chemical profile of this chemically stratified compact object. The trapping patterns of these low-radial-order $g$-modes can be visualized by observing the deviations in the mean period spacing, as well as in the observed rotational splittings, both of which are affected by structural changes in the star. We show in Figure~\ref{fig:splitsdipoles} deviations in constant period spacing and rotational splitting for the $\ell=1$ modes present in \tar. We show the same analysis for the $\ell=2$ modes in Figure~\ref{fig:splitsquads}.

Fitting a straight line to the consecutive $m=0$ modes provides a measure of the mean period spacing, which is sensitive to the overall mass of the star. For the $\ell=1$ modes we observe $\Delta \Pi_1 = 39.9\pm2.6$~s (the uncertainties are computing from the residuals between the observed periods and the linear fit). Similarly, for the $\ell=2$ modes we observe $\Delta \Pi_2 = 22.1\pm2.0$~s. The ratio expected for $\Delta \Pi_1 / \Delta \Pi_2$ from theory is $\sqrt{3}$, and we find $1.80\pm0.18$, within expectations. While neither value puts strong constraints on the stellar parameters, the relatively large period spacings suggest a relatively low-mass DBV, using the models published in \citet{Bognar14}: $<0.57$~\msun\ if \teff $>30{,}000$~K, as suggested by spectroscopy.


It is very interesting that the deviations from the mean period spacing in the $\ell=1$ modes are correlated with the differences in splittings, shown in the bottom panels of Figure~\ref{fig:splitsdipoles}. Both effects result from mode trapping. \citet{Kawaler99} show that this may be an indication of $\Omega(r)$ decreasing with $r$ (see their Figure 6). The differences appear anti-correlated for the $\ell=2$ modes shown in Figure~\ref{fig:splitsquads}.

Finally, we have explored any asymmetry in the frequency splittings that could be caused by magnetic fields \citep{Jones89}. Since we cannot see the displacement relative to the $m=0$ component in the $\ell=2$ modes, we only use the $\ell=1$ modes for this analysis\footnote{There is an interesting frequency asymmetry in the retrograde vs. prograde modes of both complete $\ell=2$ modes, $f_2$ and $f_5$. Both show significantly larger splittings between the $m=-2$ to $m=-1$ components, compared to $m=+1$ to $m=+2$.}. The frequency splittings for both $f_1$ and $f_3$ are symmetric to within the uncertainties (0.017~\muhz\ and 0.020~\muhz, respectively). Following the method and the scaling factor detailed in \citet{Winget94}, the frequency symmetry constrains the dipolar magnetic field of \tar\ to be $<10$~kG.

\section{Discussion and Conclusions}
\label{sec:conclusions}

The detection of pulsations and rotational modulation in \tar\ is exceptional for several reasons.

First, observable pulsations in such a hot DBV strongly challenges theoretical predictions of the blue edge of the DBV instability strip given current estimates of convective efficiency in DB white dwarfs \citep{Bergeron11}. The top 20 best-fit solutions found in our preliminary asteroseismic analysis agree within 2$\sigma$ with the spectroscopically inferred atmospheric parameters, found from the detailed study of \citet{Dufour10} to be \teff\ $=31{,}300\pm500$~K and \logg\ $=7.8\pm0.1$~cgs. The most recent parameters reported by \citet{Bergeron11} agree with this determination: \teff\ $=31{,}040\pm1060$~K and \logg\ $=7.83\pm0.06$. Non-adiabatic calculations do not predict pulsating DB white dwarfs with \teff\ $>29{,}000$~K \citep{Dupret08}, and trace amounts of hydrogen restrict rather than expand the instability strip \citep{Beauchamp99}.

The relatively low-amplitude pulsations observed in \tar\ demonstrate that many white dwarfs that have been observed not to pulsate, mostly from the ground, may indeed vary but at amplitudes below historical detection limits.

Additionally, \tar\ is the first pulsating white dwarf known with a photometric signal corresponding to the surface rotation period. We infer that the strongest signal originating from \tar\ corresponds to a surface rotation period of $10.17404\pm0.00067$\,hr. A folded light curve suggests this modulation arises from a hot spot on the white dwarf that is stable in amplitude and phase over the entire 78.7-d \ktwo\ light curve, to within the uncertainties. We speculate that this photometric signal is caused by redistributed flux from the ultraviolet by an opacity source localized on the surface, perhaps by a weak magnetic field. However, the symmetry in the observed pulsation splittings constrains the dipole field strength to be $<10$~kG.

We detect 11 independent pulsation modes, the majority of which have solid mode identifications and corroborate the surface rotation period. Both the lack of observed $\ell=2,m=0$ pulsation modes and spot modeling suggest we observe \tar\ at an inclination of $\sim$60 degrees. Importantly, the observed rotational splittings in these modes are not identical, since the modes are confined to different depths within the star as a result of mode-trapping effects. As such, the pulsations provide an exceptional laboratory to test the radial differential rotation and internal compositional stratification of a highly evolved stellar remnant.

Our preliminary asteroseismic analysis using a coarse grid of adiabatic models has not yielded a reliable global minimum, which is necessary to compute a full inversion to measure the radial rotation profile. We will undertake a more detailed asteroseismic analysis in a future work, and especially look forward to modeling efforts of \tar\ using the static, parametrized equilibrium structures described in \citet{Giammichele16b}.

Finally, we note that such a hot DBV is also an exquisite laboratory for plasmon neutrino production: we expect its neutrino luminosity to exceed its photon luminosity by up to 30\% \citep{Winget04}. Monitoring the rate of period change of the pulsations of \tar\ over several years will allow us to infer this excess cooling from neutrino emission. All 26 of the pulsation frequencies in \tar\ are extremely stable in phase, which is unique among hot DBVs observed over long enough baselines (e.g., \citealt{Dalessio13,Zong16}). The highest-amplitude pulsations in \tar\ all currently have limits of $dP/dt < 4 \times 10^{-10}$ s s$^{-1}$ with just this 78.7-d {\em K2} dataset. While the low amplitudes require multiple nights of observations for ground-based follow-up, the \ktwo\ data provides the perfect anchor for such a long-term project.

\acknowledgments

Support for this work was provided by NASA through Hubble Fellowship grant \#HST-HF2-51357.001-A, awarded by the Space Telescope Science Institute, which is operated by the Association of Universities for Research in Astronomy, Incorporated, under NASA contract NAS5-26555. S.D.K. acknowledges support through the K2 GO program via grant 14-K2GO2\_2-0001. J.C.C. acknowledge support from the National Science Foundation,
under award AST-1413001. This paper includes data collected by the \kep\ mission. Funding for the \kep\ mission is provided by the NASA Science Mission directorate.

{\it Facilities:} \facility{Kepler}

\end{document}